\documentclass{svjour2}                    
\smartqed  
\usepackage{graphicx}
\journalname{Journal of Statistical Physics}
\begin{document}

\title{
Autonomous Brownian motor driven by nonadiabatic variation of 
internal parameters
}


\author{A. V. Plyukhin
}


\institute{
\at              
Department of Mathematics, Saint Anselm College, Manchester, NH, USA\\
              \email{aplyukhin@anselm.edu}           
}

\date{Received: date / Accepted: date}

\maketitle

\begin{abstract}
We discuss an autonomous  motor based on a Brownian particle 
driven from thermal equilibrium  
by periodic in time variation of the internal
potential through which the particle interacts with molecules of the 
surrounding thermal bath. We demonstrate for such a motor the absence of a 
linear response regime: The average driving force and drift velocity 
are shown to be  
quadratic in both the frequency and amplitude of the variation. 
The adiabatic approximation
(of an infinitely slow variation) and the leading correction to it  
(linear in the variation's frequency) both lead to zero drift and are insufficient
to describe the motor's operation.
\end{abstract}

\section{Introduction}

Consider a Brownian particle moving  in one dimension and  
interacting with molecules from the left and  right  
through microscopic potentials $U_l$ and $U_r$ 
which are of different shapes and/or ranges.
Contrary to uncultivated intuition (but in agreement with thermodynamics), 
in thermal equilibrium such 
intrinsic  microscopic asymmetry does 
not cause a net drift of the particle:
Although a molecule, say, from the right interacts with the particle
via a stronger force than a symmetrically positioned molecule 
from the left,  
the average forces exerted on the two
sides of the particle,  
calculated with the equilibrium Boltzmann distribution,
have exactly the same magnitudes and completely compensate 
each other. On a deeper level, 
this cancellation is  ensured 
by detailed balance symmetry~\cite{Kampen} which may (or may not)  be broken  
when the system is out of equilibrium. 
Now suppose that  the particle is active in the sense that it is equipped 
with an internal mechanism which 
modulates the static potentials $U_l$ and $U_r$
in a periodic-in-time manner
\begin{eqnarray}
U_\alpha(t)=\xi_\alpha(t)\, U_\alpha,
\label{modulation1}
\end{eqnarray}
e.g. 
with harmonic modulation functions  
\begin {eqnarray}
\xi_\alpha(t)=1+a_\alpha\sin \omega_\alpha t
\label{modulation2}
\end{eqnarray} 
with amplitudes $0<a_\alpha<1$.
Here and below  the subscript  $\alpha=\{l,r\}$ refers to the left and 
right sides of the particle. 
Such modulation may result from  periodic conformational 
changes of internal degrees of
freedom,  which are very common for variety of biological macromolecules
like proteins, ribosomes, and  viruses.
Now when the particle is out of equilibrium,
one may reasonably expect that the asymmetry of static potentials $U_\alpha$, 
and/or the modulation  amplitude $a_\alpha$ and frequencies $\omega_\alpha$   
may result in the particle acquiring a nonzero average velocity 
and serving as a Brownian motor. 
As for many other machines operating
under non-equilibrium conditions, the explicit evaluation
of the drift force and velocity  is a non-trivial problem, 
and even the direction of the drift may be not easy to guess. 
There are also a number of  peculiar aspects of this model that we believe  
make it worthy to discuss.

Unlike many mesoscopic  machines driven from equilibrium 
by {\it external} means (e.g., due to a contact with  thermal 
baths of different temperatures, periodically in space and time modulated
temperature, external potential, light, etc.),  
our  motor is autonomous and resists the thermalization 
by means of the {\it internal}  mechanism. 
Among other types of  autonomous motors studied in recent years are 
granular Brownian systems~\cite{granular}, chemically powered motors 
driven by asymmetric catalytic activity~\cite{chem}, and Brownian information 
machines~\cite{info}. An autonomous 
active motor with an internal anchoring mechanism was studied in~\cite{Mecke}.

Driven by the oscillation of a {\it microscopic} parameter(s), our motor 
cannot be  described within empirical approaches, e.g. those  
based on the standard  Langevin or Fokker-Planck equations
with time dependent external parameters~\cite{Reimann_rev1,Reimann_rev2}
or modified  with the energy 
depot (``negative friction'') terms~\cite{activeBM}. Only a few microscopic models
of rectified Brownian motion are presented so far in literature, and most of them 
concern the regime beyond the weak coupling to the thermal bath.    
In such  cases, 
the 
Fokker-Planck~\cite{Gruber,Broeck_piston,Munakata,Broeck_motor1,Broeck_motor2,Broeck_motor3,Plyukhin_Froese}
and  Langevin~\cite{Plyukhin_Froese,Plyukhin_piston} equations 
involve additional terms (of higher orders in a weak coupling
parameter), and must 
be supplemented  with additional fluctuation-dissipation 
relations which cannot be established
phenomenologically.
In contrast, for the present model  
a systematic driving force will be shown to emerge already in the leading order 
in the weak coupling parameter. Still, an explicit expression
of the driving force cannot be constructed empirically and requires
a microscopic evaluation.

One interesting property of the model is the absence of the linear response
regime with respect to the modulation frequencies $\omega$ 
considered as  a perturbation parameter. 
For low $\omega$, the driving force and drift
velocity  depend on frequency  as $\omega^2$.
Neither the adiabatic approximation (asymptotically slow modulation), nor
the leading perturbational correction to it  are
sufficient to account for directional motion of the particle. 
A similar behavior  was found previously for Brownian ratchets driven by   
temperature 
oscillations~\cite{Reimann_rev1,Reimann_rev2,Reimann_motor}.
Other systems driven from equilibrium   by 
variation of external parameters may show also 
qualitatively different scenarios ranging from  directional transport driven by
adiabatically slow  variations~\cite{Parrondo,Astumian} to the absence 
of  net transport beyond the adiabatic regime~\cite{Jar,Cher}. 
Whether such diversity can also be observed in the family of autonomous machines
like ours is yet to be explored.

It might be relevant  to note that the  absence of the linear response 
regime was previously observed  also 
for  motors  with non-varying parameters but
driven by the coupling  with two baths with non-equal  
temperatures~\cite{Broeck_motor1,Broeck_motor2,Broeck_motor3}.
In that case,  the prediction that the drift velocity must be an even function
of a perturbation parameter (the temperature difference) can be envisaged 
from simple symmetry arguments. 
No similar arguments are apparently  available for the present model.

Another remarkable feature of the 
model, which also has counterparts among  motors driven by 
variation of external parameters~\cite{Reimann_motor,Reimann_motor2},  
is a  nonmonotonic dependence and - for certain regimes - the sign reversal
of the drift velocity as a function of  $\omega$ (see Fig. 3 below).  
However, for our model   
such behavior is probably of only academic interest since it 
occurs when the period of modulation $\omega^{-1}$ is
unrealistically short - of order or shorter than the collision time $\tau$.
Since the latter is usually the shortest characteristic time,  
we shall restrict our theoretical discussion to the low frequency limit
$\omega\,\tau\ll1$. Numerical simulation will be used
to confirm theoretical predictions and to extend the results  beyond the 
low frequency domain.

\section{Model and simulation}
\label{sec:1}

We consider a Brownian particle of mass $M$ immersed in  the 
thermal bath of temperature $T$ comprised of 
ideal gas molecules of mass $m\ll M$. 
Velocities of molecules before collisions with the particle are
distributed with the  Maxwell distribution 
\begin{eqnarray}
f_M(v)=\frac{1}{\sqrt{2\pi}\,v_T}\,\exp\left\{-\frac{1}{2}\,
\left(\frac{v}{v_T}\right)^2\right\},
\end{eqnarray}
where $v_T=\sqrt{k_B\,T/m}$ is a thermal (mean-squared) velocity
of a molecule. The motion of the particle and molecules of the
bath is assumed to be one-dimensional.

\begin{figure}[htb]
\centerline{\includegraphics[height=5cm]{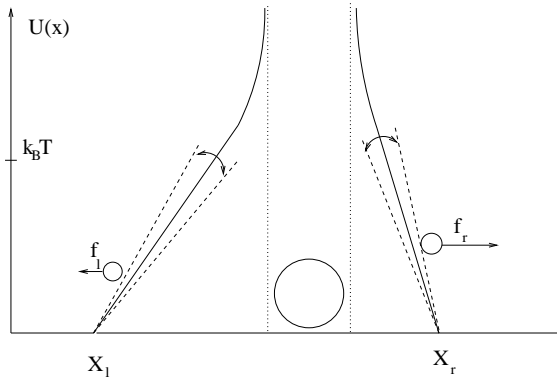}}
\caption{The solid line is 
the potential energy for the interaction of the motor 
(large circle) in the passive regime  
and  a molecule of the bath (small circles)
as a function of position of the latter. When a molecule is
inside the left (right) interaction zone $x>X_l$ ($x<X_r$), it experiences
a constant repulsive force of the amplitude $f_l$ ($f_r$). 
It is assumed that temperature is
sufficiently low, so that only a negligible number of molecules
with energy much higher than $k_BT$ experience 
the upper nonlinear part of the potential. Dashed lines represent a periodic
in time modulation of the potential for the active regime.
}
\label{fig1}
\end{figure}

Dynamics of the particle's internal 
degrees of freedom, are governed by a certain internal built-in mechanism whose 
specific design is immaterial for our purposes. When this mechanism 
is turned off the particle is ``passive'', i.e. behaves as a 
conventional Brownian particle interacting with molecules of the bath with
time-independent forces. 
In order to facilitate analytic calculations,
we assume that in the passive regime  the bath
molecules, unless their energy is too high, interact with the particle  
with constant repulsive forces of a finite range,
see Fig. 1.  Namely, 
the potential energy of interaction of the particle
with bath molecules on its left reads
\begin{eqnarray}
U_l=
f_l \,\sum_i(x_i-X_l)\,\theta(x_i-X_l).
\label{Ul}
\end{eqnarray}
Here $x_i$ are coordinates of bath molecules, 
$f_l$ is a positive constant of the force dimension, 
$\theta(x)$ is the step-function. The coordinate $X_l$
is associated with the left side of the particle and  
determines the boundary of the interaction zone for molecules coming from the
left: molecules outside the interaction zone, $x_i<X_l$, do not interact with
the particle, while  every  molecule inside the zone, $x_i>X_l$, 
exerts on the particle the same force $f_l$.
Similarly, the potential energy of interaction  with molecules
on the right of the particles is  
\begin{eqnarray}
U_r=
-f_r\sum_j(x_j-X_r)\,\theta(X_r-x_j), 
\label{Ur}
\end{eqnarray}
where the positive constant $f_r$ is the amplitude of the constant repulsive 
force exerted on the particle by a bath  molecule from the right 
when the latter is in the right-hand side interaction zone, $x_j<X_r$. 

To avoid complications related to the overlapping
of left and right interaction zones, one can assume that 
closer to the
particle's core the linear potential is replaced by a sharper (diverging) one, 
see Fig. 1. This, however, is of no consequence as soon as 
the crossover from the linear to nonlinear potentials occurs 
at the energy much higher than $k_BT$: only a negligible fraction of molecules
with velocities $v\gg v_T$ would feel the nonlinear part of the potential.

For this model the fluctuating Langevin force $F(t)$ and its correlations can be
readily evaluated analytically (see section 3 in~\cite{Plyukhin_motor}). 
To the lowest order
in the mass ratio parameter $\lambda=\sqrt{m/M}$ and for the time scale 
much longer than the collision time (defined by Eq. (\ref{tau_T}) below) 
the Langevin equation for the particle's velocity $V$ has the standard form
\begin{eqnarray}
M\,\frac{d\, V}{dt}=-\gamma \, V+ F(t).
\label{LE}
\end{eqnarray} 
A possible asymmetry of the microscopic forces, $f_l\ne f_r$,  
does not show up in this equation.
Although the correlation function of the Langevin force 
$\langle F(0) F(t)\rangle$ does depends
on $f_l$ and $f_r$, this dependence disappears after the integration over
time. As a result, 
the damping coefficient $\gamma$  does not depend on parameters
of microscopic dynamics and  takes the form 
\begin{eqnarray}
\gamma=\frac{1}{k_B T}\,\int_0^\infty \langle F(0)F(t)\rangle\, dt=
4\sqrt{\frac{2}{\pi}}\, n\, m\, v_T,
\label{gamma}
\end{eqnarray} 
where  $n$ is the concentration of bath molecules. 
Also, if one writes $F$
as  a sum of forces on the left and right sides of the particle, $F=F_l+F_r$,
one can show that
\begin{eqnarray}
\langle F_l\rangle=-\langle F_r\rangle=n\, k_BT,
\label{passive_force}
\end{eqnarray}
so the net Langevin  force is zero-centered, $\langle F(t)\rangle=0$.
In accord with thermodynamics, the  microscopic asymmetry does not bias 
equilibrium Brownian motion.

Our goal is to generalize the above microscopic  model of passive 
Brownian motion
for the case when the particle is ``active'', that is capable to variate the
potentials $U_l$ and $U_r$ according to Eq. (\ref{modulation1}).
For the active particle
the potential energy of interaction with the bath is
 given by expressions similar to that for the passive regime
\begin{eqnarray}
U_l(t)&=&f_l (t)\,\sum_i(x_i-X_l)\,\theta(x_i-X_l)\nonumber\\
U_r(t)&=&-f_r(t)\sum_j(x_j-X_r)\,\theta(X_r-x_j), 
\label{U}
\end{eqnarray}
in which the
static force magnitudes  $f_l$ and $f_r$ are now 
replaced by time-dependent ones,
\begin{eqnarray}
f_\alpha(t)=\xi_\alpha(t)\, f_\alpha,
\label{force_t}
\end{eqnarray}
where $\xi_\alpha(t)$ is given by (\ref{modulation2}),
$\xi_\alpha(t)=1+a_\alpha\sin\omega_\alpha t$  (to minimize 
the number of parameters we assume no phase shift).  
Clearly, such a variation requires an external input of energy, which is not
explicitly reflected in the model.

\begin{figure}[htb]
\centerline{\includegraphics[scale=0.65]{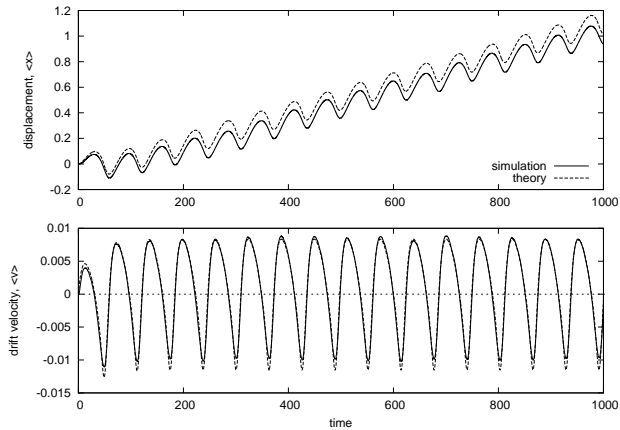}}
\caption{Average displacement $\langle x(t)\rangle$ and
velocity  $\langle v(t)\rangle$ of the motor as functions of time for 
regime (\ref{A}) with parameters  
$\omega_r=\omega_l=0.1$, $a_r=a_l=0.5$, and $\delta=f_r/f_l=3$.
Numerical experiment data (solid lines) are averaged over about 
$2\cdot 10^5$ trajectories. Theoretical curves (dashed lines) are solutions
of the Langevin equation (\ref{LE_dimless_1}) which corresponds to 
the average fluctuating force 
$\langle F(t)\rangle$ given by Eq. (\ref{F_A}). 
Units are defined by Eqs. (\ref{units}).    
}
\label{fig2}
\end{figure}

The modulation frequencies
$\omega_\alpha$ will be assumed to be small compared to the characteristic 
collision time, while the amplitudes $a_\alpha$ are not necessarily small and
may take any values from the interval  $(0, 1)$. The values $a_\alpha=1$,
though   not altogether meaningless (may correspond to  a permeable 
particle, see~\cite{Plyukhin_motor}), are not  considered.
The variation of the particle-bath  interaction supports the particle 
in a nonequilibrium state, which is the first condition of the drift. 
The second condition
- the break of the spatio-temporal symmetry - can be arranged by
assigning different sets of values $\{f_\alpha, a_\alpha,  \omega_\alpha\}$ for
the left and right sides of the particle.

We study this model both theoretically and using numerical simulation. The
latter is performed for the
mass ratio $\lambda^2=m/M=0.01$
with the standard molecular dynamics scheme with the
only difference that instead of using periodic boundary conditions,  we
generate in the beginning of each simulation run a very large thermal bath of
noninteractive molecules (see~\cite{Plyukhin_motor} for details). 
The simulation shows that for unequal sets
$\{f_l, a_l,  \omega_l\}$ and 
$\{f_r, a_r,\omega_r\}$ the particle develops non-stationary drift velocity, 
whose direction and value depend on  $f_\alpha, a_\alpha,\omega_\alpha$
in a rather subtle way.

\begin{figure}[htb]
\centerline{\includegraphics[scale=0.6]{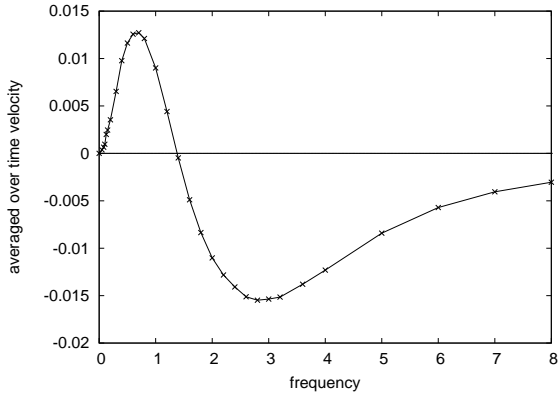}}
\caption{ Drift velocity of the motor averaged over time as a function of
the modulation frequency $\omega$ for
regime (\ref{A})  with parameters  $a_r=a_l=0.5$, and $\delta=f_r/f_l=3$.
Each experimental point corresponds to an ensemble-averaged trajectory 
similar to that in Fig. 2. 
}
\label{fig3}
\end{figure}

As a showcase example let us consider the case 
 when asymmetry is due to non-equal magnitudes of the static (passive) 
forces, $f_l\ne f_r$. Specifically, 
suppose that the  particle, while in the passive regime, 
interacts with a  molecules on the right
via a stronger static force than with a molecule from the left (as in Fig. 1), 
while the modulation parameters 
for the left and right sides are the same,
\begin{eqnarray}
f_l<f_r,\quad \omega_l=\omega_r=\omega, \quad a_l=a_r=a. 
\label{A}
\end{eqnarray}
For this case, when  the modulation  frequency is sufficiently low,
the simulation shows that 
the particle drifts to the right, that is in the direction of the 
steeper slope of
the internal potential,  see Fig. 2.  Perhaps somewhat counter-intuitive, this 
behavior is in agreement with a theory which will be developed in 
sections to follow.  Note  that for the given values of parameters 
the drift is rather small and becomes visible  
only after taking average over many particle's trajectories.  
A single trajectory looks like a random path of a passive
Brownian particle and shows no visible  bias. 
All experimental curves in this paper represent data averaged over 
about $10^5$ trajectories.

Another interesting feature suggested by  simulation data is  that
for low $\omega$ the drift velocity increases with frequency  as
$\omega^2$. This can be interpreted as the absence of the linear response
regime with respect to $\omega$ as a perturbation.
As frequency is getting higher, the frequency dependence of the drift velocity
becomes nonmonotonic, and the inversion of the drift direction occurs,
see Fig. 3.

According to (\ref{force_t}),  
molecules in the left and right interaction zones of the motor 
experience  the forces of magnitudes
\begin{eqnarray}
f_\alpha(t)=f_\alpha+f_\alpha\,a_\alpha \,\sin\omega_\alpha t.
\end{eqnarray} 
One observes  that for the regime defined by Eq. (\ref{A}) 
both static and time-dependent parts of these expressions are different
for the left and right sides of the particle.
It is natural to ask if the drift still occurs
when asymmetry affects only the static parts, 
\begin{eqnarray}
f_l < f_r\quad 
f_la_l=f_r\,a_r, \quad \omega_l=\omega_r.
\label{B}
\end{eqnarray}
or only dynamic parts, e.g. 
\begin{eqnarray}
f_l = f_r\quad 
a_l<\,a_r, \quad 
\omega_l=\omega_r.
\label{C}
\end{eqnarray}
The simulation shows the drift in both these cases too, 
but with no sign reversal
and in opposite directions, see Fig. 4. This suggests that the reversal of the
drift direction for regime (\ref{A}) can be interpreted as a 
result of the  interplay or interference of two regimes described by Eqs.
(\ref{B}) and (\ref{C}). 

\begin{figure}[htb]
\centerline{\includegraphics[scale=0.6]{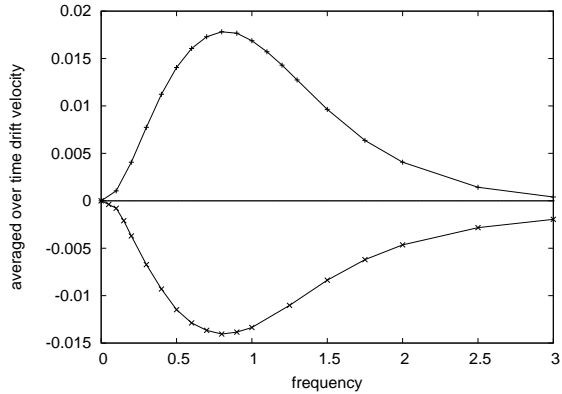}}
\caption{Drift velocity of the motor averaged over time as a function of
the modulation frequency $\omega$. The upper curve corresponds to the regime
(\ref{B}) (asymmetric static part of interaction) with 
 $f_r/f_l=3$ and $f_l\,a_l=f_r\,a_l=0.5$. The bottom curve
is for regime (\ref{C}) (asymmetric dynamical part of interaction) 
with $a_r=0.25$ and $a_r=0.5$.}
\label{fig4}
\end{figure}

Another interesting regime of the motor's operation is
when the drift is  induced by unequal  modulation 
frequencies  (one frequency may be zero), 
while the pair of other parameters is the same for both sides of
the particle,  
\begin{eqnarray}
f_l = f_r\quad 
a_l=a_r, \quad 
\omega_l\ne\omega_r.
\label{D}
\end{eqnarray}
In this case,  if  one or both frequencies   are  low 
(much shorter than the collision time), the particle drifts in the
direction of the side with lower modulation frequency 
(two upper curves in Fig. 5). 
On the other hand, 
if both frequencies are high, the particle systematically moves in the
direction of the side with a higher frequency (a bottom curve  in Fig. 5).

In the following sections we shall focus on a theoretical description of 
the model.
We shall  not try to cover the whole rich
phenomenology which the model shows in simulation, but rather
restrict ourselves to developing a perturbation approach  for the low frequency
domain.

\begin{figure}[htb]
\centerline{\includegraphics[scale=0.6]{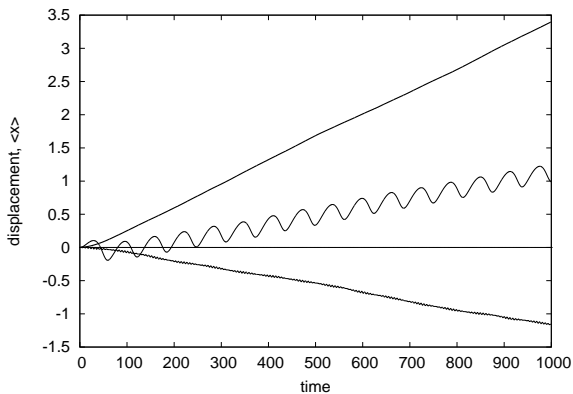}}
\caption
{Average displacement $\langle x(t)\rangle$ of the motor as a function
of time for regime (\ref{D}) when asymmetry is due to unequal
modulation frequency, $\omega_l\ne\omega_r$. The upper line is for
$\omega_l=2.5$ and $\omega_r=0$, the middle line is for $\omega_l=0.1$ and 
$\omega_r=0$, and  the bottom  line is for $\omega_l=1.1$ and 
$\omega_r=1$. For all lines the modulations amplitudes are the same
$a_l=a_r=0.5$. The oscillatory character of the upper line is imperceptible
on the figure's scale.  
} 
\label{fig5}
\end{figure}

\section{Theory: basic relations}
We assume that in the active regime the particle is still described by the
Langevin equation (\ref{LE}) with the damping coefficient approximately the
same as for the passive regime, i.e. given by  (\ref{gamma}), 
but with a non-zero centered fluctuating force, 
$\langle F(t)\rangle\ne 0$. The goal is to evaluate
microscopically $\langle F(t)\rangle$ and  then, solving the averaged
Langevin equation
\begin{eqnarray}
M\, \frac{d}{dt}\,\langle V(t)\rangle&=& -\gamma \,\langle V(t)\rangle +\langle
F(t)\rangle,
\label{LE_V}
\end{eqnarray} 
find the average  velocity 
$\langle V(t)\rangle$ and trajectory $\langle X(t)\rangle$ of the motor.

An obvious drawback of this approach is that it neglects the influence of  the
active part of the fluctuating force $F(t)$ on the damping coefficient
$\gamma$. Since the two quantities are related by a
fluctuation-dissipation relation, this approximation cannot be entirely
consistent. Yet  it is clear that for sufficiently small $\omega_\alpha$
or/and $a_\alpha$ an ``active''  correction  to the
dissipation constant is small compared to the value of the latter for the
passive regime, and thus should produce a little effect. We shall see that 
the comparison of the theory with simulation results supports this intuition.

As known from the microscopic theory of Brownian motion, the Langevin equation 
(\ref{LE})  corresponds to the lowest order approximation in the mass ratio
$m/M$,  in which case the fluctuating force $F(t)$ can be evaluated
neglecting the particle's motion. Setting $F=F_l+F_r$,  let us 
consider the force $F_l$ exerted on the particle,  fixed in space,  
by molecules coming from the left. 
For the linear potential  (\ref{U}),
each molecule in the left interaction zone  $x>X_L$  
exerts on the particle the same 
time dependent force $f_l(t)$.
Then the total force on the left side is given simply by the product
\begin{eqnarray}
F_l(t)=f_l(t)\,N_l(t),
\label{F1}
\end{eqnarray}
where the $N_l(t)$ is the number of molecules in the left  interaction 
zone at a given time,
\begin{eqnarray}
N_l(t)=\int_{-\infty}^\infty dv \int_{X_l}^\infty  dx\, f(x,v,t).
\label{N1}
\end{eqnarray}
Here   $f(x,v,t)=\sum_i\delta(x-x_i)\delta(v-v_i)$ 
is the microscopic density of molecules in the position and
velocity space. For simplicity we extended the right border of the left
interaction zone to infinity.  
For low enough temperature the effect of such approximation is negligible. 

It is convenient to define the collision  time $\tau_l(v,t)$ 
as a time spent in the interaction zone $x>X_l$ by a molecule which
enters the zone with velocity $v$ and {\it  leaves} the zone at time $t$. 
The collision time $\tau_r(v,t)$ for molecules hitting the particle's right side
is defined in a similar way. 
The dependence of $\tau_\alpha(v,t)$ on time reflects the dynamical nature of 
the interaction potentials (\ref{U}).  
Of course, for the passive regime the  collision times  
depends on molecule's pre-collision velocity only. 

\begin{figure}[htb]
\centerline{\includegraphics[height=4.5cm]{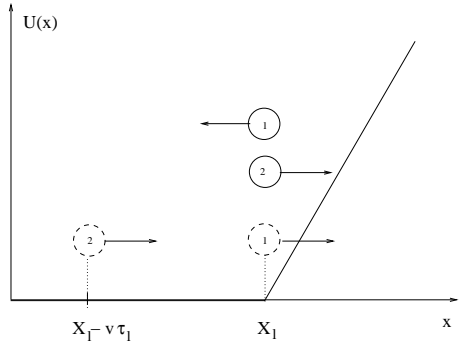}}
\caption{Solid line circles: at time $t$   
molecule two is just before the collision entering the left interaction zone
$x>X_l$, and molecules one is just after the collision leaving 
the interaction zone. Dashed line circles show same molecules at time
$t-\tau_l$.  
}
\label{fig6}
\end{figure}

Consider two molecules with  the same pre-collision  velocity $v>0$ 
at the border of  the interaction zone $X_L$ at time $t$.  
Molecule one is just after the collision,  departing the interaction zone,
\begin{eqnarray}
x_1(t)=X_l, \qquad v_1(t)<0,
\end{eqnarray}
while molecule two is just before the  collision,  
entering the interaction zone,  
\begin{eqnarray}
x_2(t)=X_l, \qquad v_2(t)=v>0,
\end{eqnarray}
see Fig. 6.
According to the definition of $\tau_\alpha(v,t)$, 
at earlier time $t-\tau_l(v,t)$ both molecules had velocity $v>0$ and 
coordinates
\begin{eqnarray}
x_1(t-\tau_l)=X_l, \qquad x_2(t-\tau_l)=X_l-v\,\tau_l.
\end{eqnarray}
For the low modulation frequency it is reasonable to 
assume that molecules with the same initial velocity do not bypass
each other in the interaction zone. 
Then it is clear that
{\it all}  molecules with a pre-collision velocity $v>0$ which are 
in the interaction zone at time $t$, at the time $t-\tau_l$
were within the interval 
\begin{eqnarray}
[x_1(t-\tau_l), x_2(t-\tau_l)]=[X_l-v\,\tau_l, X_l]      
\end{eqnarray}
and moving to the right with velocity $v$, see Fig. 6. 
Thus for the number of molecules
in the interaction zone at a given time, instead of (\ref{N1}) one can write 
\begin{eqnarray}
N_l(t)=\int_{0}^\infty dv \int_{X_l-v\tau_l}^{X_l} dx\, f(x,v,t-\tau_l).
\label{N2}
\end{eqnarray}
The advantage of this form is that it involves integration over
coordinates and velocities of molecules before collisions with the particle, 
in which case the microscopic density is simply the Maxwell distribution 
 $f_M(v)$ multiplied by the average concentration of bath molecules $n$, 
\begin{eqnarray}
\langle f(x,v,t)\rangle= n\, f_M(v). 
\end{eqnarray}
Then from (\ref{N2}) for the average number of molecules in the interaction
zone one obtains
\begin{eqnarray}
\langle N_l(t)\rangle=n\,\int_{0}^\infty dv\,f_M(v)\,v\,\tau_l(v,t), 
\label{N3}
\end{eqnarray}
and  the average force on the left side of the particle 
$\langle F_l(t)\rangle=f_l(t)\,\langle N_l(t)\rangle$
takes the form
\begin{eqnarray}
\langle F_l(t)\rangle=n\,f_l(t)\,\int_{0}^\infty dv\,f_M(v)\,v\,\tau_l(v,t). 
\label{F2L}
\end{eqnarray}
Similarly, the average force on the right side of the particle 
is
\begin{eqnarray}
\langle F_r(t)\rangle=-n\, f_r(t)\,\int_{-\infty}^0 dv\,f_M(v)\,v\,\tau_r(v,t). 
\label{F2R}
\end{eqnarray}

To advance further, we need  to evaluate 
collision times $\tau_\alpha(v,t)$ for the active regime,
which will be the focus of the next section. Meanwhile, it is
instructive to consider how the general expressions (\ref{F2L}) and
(\ref{F2R}) work for the passive regime, when   
$f_\alpha(t)=f_\alpha=const$. In this case the collision times 
do not depend on time and we shall denote them as $\tau_\alpha^0(v)$.
Consider a molecule which  
enters, say,  the left interaction zone at the moment $t=0$ with the
initial velocity $v>0$. While the molecule is still in the  
zone its velocity evolves as
\begin{eqnarray}
v(t)=v-\frac{f_l}{m}\, t.
\end{eqnarray}    
The collision time 
$\tau_l^0$ corresponds to the moment when the molecule  
leaves the zone with velocity $-v$. The equation $v(\tau_l^0)=-v$ gives
\begin{eqnarray}
\tau_l^0(v)=\frac{2mv}{f_l}.
\label{tau_passive_l}
\end{eqnarray}
Similarly, the right side collision  time is 
\begin{eqnarray}
\tau_r^0(v)=\frac{2mv}{f_r}.
\label{tau_passive_r}
\end{eqnarray}
Substitution of these expression into Eqs.(\ref{F2L}) and (\ref{F2R})
gives the result (\ref{passive_force}) of the elementary kinetic theory
\begin{eqnarray}
\langle F_l\rangle=-\langle F_r\rangle\equiv F_0=n\,m\,\langle v^2\rangle=
n\,k_B\,T.
\label{F_eq}
\end{eqnarray}

As expected, for the passive regime the asymmetry $f_l\ne f_r$ does not 
induce  a net driving  force on the particle. We can now interpret this no-go
result as follows:
Each molecule in  the $\alpha$-th interaction zone  pushes the particle 
with a force of magnitude $f_\alpha$, 
but the average zone's population $\langle N_\alpha\rangle$, according to (\ref{N3}),  
is linear in the  collision time $\tau_\alpha$, which in turn is inversely
proportional to $f_\alpha$, $\tau_\alpha\sim 1/f_\alpha$. 
As a result, for the average force 
$\langle F_\alpha\rangle\sim f_\alpha\,\langle N_\alpha\rangle$ the 
dependence  on $f_\alpha$ is canceled out.

\section{Collision time for active regime}
Above we defined the collision time $\tau_\alpha(v,t)$ as functions of 
the pre-collision velocity $v$  and after-collision exit time $t$,
which corresponds to the moment  
when a molecule is {\it leaving} the interaction zone. 
With these functions one can express the forces $\langle F_\alpha(t)\rangle$ 
on two particle's sides in simple forms  (\ref{F2L}) and (\ref{F2R}).
The disadvantage  of functions $\tau_\alpha(v,t)$ is that they are difficult
to  evaluate directly.  Let us define  
collision times $\tau_\alpha^*(v,t)$ whose 
first argument is still velocity before the collision, while the time
argument refers
now to the moment when a molecule {\it enters}  the interaction zone.
Since both arguments of $\tau_\alpha^*(v,t)$
refer to the same pre-collision moment, these functions are easier 
to evaluate.  
The two functions $\tau_\alpha(v,t)$ and
$\tau_\alpha^*(v,t)$ are related by the equation
\begin{eqnarray}
\tau(t)=\tau^*\big(t-\tau(t)\big).
\label{taus}
\end{eqnarray}  
Here and for the bulk of this section  we omit for brevity 
the argument $v$ which is assumed to be 
the same  for all quantities, and  
also suppress the left/right index $\alpha$.  
We shall restore $\alpha$ in the final expression for $\tau_\alpha$.

In order to find an explicit expression for $\tau$
in terms of $\tau^*$ we approximate the right hand side of Eq. (\ref{taus}) 
by the first three terms of the  Taylor expansion about $t$,
\begin{eqnarray}
\tau\approx\tau^*-(\tau^*)'\,\tau+\frac{1}{2}(\tau^*)''\,\tau^2,
\label{taus2}
\end{eqnarray}
where primes denote time derivatives.
Since $(\tau^*)'\sim
\omega$ and $(\tau^*)''\sim\omega^2$, the above relation is of second 
order in $\omega$. To the same order, the solution of  
Eq. (\ref{taus2}) has a form 
\begin{eqnarray}
\tau=\tau^*-(\tau^*)'\,\tau^*+\frac{1}{2}\,(\tau^*)''\,(\tau^*)^2
+[(\tau^*)']^2\,\tau^*,
\label{taus3}
\end{eqnarray}
which produces  a desirable explicit expression of $\tau$ in terms of $\tau^*$.
Our goal now is first to evaluate $\tau^*$, and 
then using (\ref{taus3}) to find $\tau$. Substitution of   
$\tau$ into Eqs. (\ref{F2L}) and (\ref{F2R}) will give us
a perturbation expression for the force on the particle to second order in $\omega$.      

Consider a molecule which enters the left interaction zone $x>X_l$ 
with velocity 
$v_0>0$ at the moment $t=t_0$. Setting for a moment $X_l=0$,  
the equation of motion 
and initial conditions read
\begin{eqnarray}
&&m\,x''(t)=-f(t)=-f\,\xi(t)\nonumber\\ 
&&x(t_0)=0, \quad x'(t_0)=v_0,
\label{eofm}
\end{eqnarray}
where $\xi(t)$ is given by (\ref{modulation2}).
The solution of the initial value problem (\ref{eofm}) 
is convenient to write as a function of time $t$ elapsed since the moment 
$t_0$ when the molecule enters the zone:
\begin{eqnarray}
\!\!\!\!\!
&&x(t)=v_0\,t-\frac{f}{2m}\,t^2 +\nonumber\\
&&\frac{f \,a}{m\omega^2}\Big(\sin\omega (t_0+t)-\sin\omega t_0-
\omega t\,\cos\omega t_0\Big).
\label{x_exact}
\end{eqnarray}
Recall that for the passive regime, the collision time $\tau^*$
can be determined
from the equation $v(\tau^*)=-v$, since the speed of a molecule 
before and after the collision is the same. That is, of course, not 
so for a time-dependent potential. For the active regime  
the collision time $\tau^*(v_0,t_0)$ should be found as 
a nonzero solution of the equation
\begin{eqnarray}
x(\tau^*)=0.
\label{eq_for_tau}
\end{eqnarray} 
We wish to find an approximate solution of this equation  to 
order $\omega^2$ (an appropriate dimensionless small parameter is introduced by 
equation (\ref{parameter}) below). 
Using in (\ref{x_exact}) the truncated expansion 
\begin{eqnarray}
&&\sin\omega(t_0+t)\approx\sin\omega t_0+\cos\omega t_0\,(\omega t)-
\frac{1}{2}\,\sin\omega t_0\,(\omega t)^2\nonumber\\
&&-\frac{1}{6}\,\cos\omega t_0 \,(\omega
t)^3+\frac{1}{24}\,\sin\omega t_0\,(\omega t)^4,
\end{eqnarray}
one obtains to order $\omega^2$
\begin{eqnarray}
\!\!\!\!\!\!
x(t)=x_{ad}(t)
-\frac{f\, a \,\omega \,\cos\omega t_0}{6\,m}\, \,t^3
+\frac{f\, a \,\omega^2\,\sin\omega t_0}{24 \,m}\,\, t^4.
\label{xapprox}
\end{eqnarray}
Here the first term is of zero order in $\omega$ 
\begin{eqnarray}
x_{ad}(t)=v_0t-\frac{f(t_0)}{2m}\, t^2,
\label{x_ad}
\end{eqnarray}
and corresponds to the adiabatic approximation
which completely neglects the change of the potential during the collision.
As we shall see below, 
the adiabatic approximation $x(t)\approx x_{ad}(t)$ is not sufficient 
to account for the motor's drift.   
The two last terms in (\ref{xapprox}) are of first and second order in
$\omega$, and take into account the dynamical nature of the potential.

Substitution of (\ref{xapprox}) into (\ref{eq_for_tau}) and resetting
$(v_0, t_0)\to (v,t)$ gives
for the collision time $\tau^*(v,t)$ a  cubic algebraic equation
which we write in the following dimensionless form:
\begin{eqnarray}
\!\!\!\!\!\!\!\!
\xi(t)\,
\left(\frac{\tau^*}{\tau^0}\right)
+\frac{1}{3} \,a\, (\omega\tau^0)\, \cos\omega t\,
\left(\frac{\tau^*}{\tau^0}\right)^2
-\frac{1}{12}\, a\, (\omega\tau^0)^2 \,
\sin\omega t\,\left(\frac{\tau^*}{\tau^0}\right)^3=1,
\label{eq_algebra}
\end{eqnarray}
where, recall,  $\tau^0=\tau^0(v)=2mv/f$ is the collision 
time for the passive regime. 
Solving this equation perturbatively to order $\omega^2$,
one obtains (see Appendix):
\begin{eqnarray}
\!\!\!\!\!\!\!\!\!
\tau^*&=&\tau^0\Bigl\{
\xi^{-1}(t)-\frac{a}{3}\,\,(\omega\tau^0)\,\,\xi^{-3}(t)\,\cos\omega t\nonumber\\
 &+&\frac{a}{12}\,\,(\omega\tau^0)^2\,\,\xi^{-4}(t)\, \sin\omega t
+\frac{2a^2}{9}\,\,(\omega\tau^0)^2\,\,\xi^{-5}(t)\, \cos^2\omega t
\Bigr\}.
\label{tau_star}
\end{eqnarray}

Recall that $\tau^*(v,t)$ is a function of a time 
when a molecule enters the interaction zone, 
while the expressions (\ref{F2L}) and (\ref{F2R})  
for the average forces involves
the collision time $\tau(v,t)$ as a function of the time when a molecule leaves the zone.
To second order in $\omega$ the relation between $\tau^*$ and $\tau$  
is given by Eq. (\ref{taus3}). Substituting (\ref{tau_star}) into
(\ref{taus3}), neglecting terms of order higher than $\omega^2$, and restoring
the right/left index $\alpha$ we obtain
\begin{eqnarray}
\!\!\!\!\!\!\!\!\!\!\!\!\!
\tau_\alpha&=&\tau^0_\alpha\,\,\Bigl\{
\xi_\alpha^{-1}(t)+\frac{2a_\alpha}{3}\,\,
(\omega_\alpha\tau^0_\alpha)\,\,\xi^{-3}_\alpha(t)\,\,\cos\omega_\alpha t\nonumber\\
 &+&\frac{a_\alpha}{4}\,\,(\omega_\alpha\tau^0)^2\,\,\xi_\alpha^{-4}(t)\,\, 
\sin\omega_\alpha t
+\frac{8a^2_\alpha}{9}\,\,(\omega_\alpha\tau^0_\alpha)^2\,\,\xi^{-5}_\alpha(t)\,\, 
\cos^2\omega_\alpha t 
\Bigr\}.
\label{tau}
\end{eqnarray}  

To zeroth order in $\omega$ this expression gives
\begin{eqnarray}
\tau_\alpha(v,t)\approx\tau_\alpha^0(v)\,\xi_\alpha^{-1}(t)=
\frac{2mv}{f_\alpha(t)}.
\label{adiabatic}
\end{eqnarray}
This  is just the expression for the collision time $\tau_\alpha^0$ for 
the passive regime, given  
Eqs. (\ref{tau_passive_l}) and (\ref{tau_passive_r}),  
in which the static force magnitudes
$f_\alpha$ are replaced by dynamic ones $f_\alpha(t)=f_\alpha\,\xi_\alpha(t)$.  
Thus the expression (\ref{adiabatic})  corresponds to 
the adiabatic approximation and, as one can check,  can be derived as a
solution of the equation $x_{ad}(\tau^*)=0$ with $x_{ad}(t)$ given by
(\ref{x_ad}).

Three last terms in the right hand side  of Eq. (\ref{tau}) describe 
corrections 
to the  adiabatic approximation up to order $\omega^2$. These terms 
behave differently being time-averaged over the period of modulation: the term
linear in $\omega $ vanishes, whereas 
the terms quadratic in $\omega$ do not 
vanish and have opposite signs.  As we shall see in the next section, it is
these two last terms which are responsible for the particle's drift.

\section{Driving force}
Substitution of Eq. (\ref{tau}) for the collision times into
Eqs. (\ref{F2L}) and (\ref{F2R}) yields  the following expression 
for the average forces on two sides of the particle:
\begin{eqnarray}
&&\langle F_\alpha(t)\rangle =\pm\, F_0\,\Bigl\{\,1+ 
c_1\,\,(\omega_\alpha\hat\tau_{\alpha})\,\,a_\alpha\,\,\xi_\alpha^{-2}(t)\,\, 
\cos\omega_\alpha t\nonumber\\
&+&c_2\,\,(\omega_\alpha\hat\tau_{\alpha})^2\,\,a_\alpha\,\,\xi_\alpha^{-3}(t)\,\, 
\sin\omega_\alpha t
+c_3\,\,(\omega_\alpha\hat\tau_{\alpha})^2\,\,a_\alpha^2\,\,\xi_\alpha^{-4}(t)\,\, 
\cos^2\omega_\alpha t\,
\Bigr\}.
\label{F_alpha}
\end{eqnarray}
Here the sign of the right hand side 
is plus  and minus  respectively for the force
on the left ($\alpha=l$) and right ($\alpha=r$)  sides of the particle,
$F_0$ is the magnitude of the force on each side for 
the passive regime given by (\ref{F_eq}), 
\begin{eqnarray}
F_0=nk_BT=n\,m\,v_T^2,
\end{eqnarray} 
$\hat\tau_\alpha$ is the collision time for the passive regime 
for a molecule with the thermal speed $v_T=\sqrt{k_BT/m}$,
\begin{eqnarray}
\hat\tau_\alpha=\tau_\alpha^0(v_T)=\frac{2\,m\,v_T}{f_\alpha},
\label{tau_T}
\end{eqnarray}
and numerical coefficients are
\begin{eqnarray}
c_1=\frac{4}{3}\,\sqrt{\frac{2}{\pi}}, \quad c_2=\frac{3}{4}, \quad 
c_3=\frac{8}{3}.
\end{eqnarray}

The above expression (\ref{F_alpha}) is  a perturbation expansion 
of the average force up to second order in $\omega$, or more precisely,  in the
dimensionless parameter
\begin{eqnarray}
\epsilon_\alpha=\omega_\alpha \hat\tau_\alpha.
\label{parameter}
\end{eqnarray} 
The approximation of zeroth order
$\langle F_\alpha(t)\rangle\approx\pm\, F_0$
coincides with the  force
for the passive regime, Eq. (\ref{F_eq}), and 
corresponds to the adiabatic approximation (\ref{adiabatic}) 
for the collision time which neglects
the variation of the potential during the collision.
As expected, in the adiabatic approximation,  asymmetry 
of the static force magnitudes $f_\alpha$  and of parameters of 
dynamic modulation $(a_\alpha, \omega_\alpha)$ do not
show up, and the   
the total average force on the particle 
$\langle F\rangle=\langle F_l\rangle+\langle F_r\rangle$ is zero.

The lowest order correction to the adiabatic approximation 
in linear in $\omega_\alpha$
(or $\epsilon_\alpha$) and  
is presented by the second term
in the right hand side of Eq.(\ref{F_alpha}). 
If one considers $\omega$ as a variable responsible for perturbation from
equilibrium, the truncation
\begin{eqnarray}
\!\!\!\!\!
\langle F_\alpha(t)\rangle \approx\pm\, F_0\,\Bigl\{\,1+ 
c_1\,\,(\omega_\alpha\hat\tau_{\alpha})\,a_\alpha\,\xi_\alpha^{-2}(t)\,\, 
\cos\omega_\alpha t
\Bigr\}
\label{F_linear}
\end{eqnarray}
can be interpreted as 
a linear response approximation.
In contrast to the adiabatic approximation, 
asymmetry of static and dynamic parameters 
is manifestly present here, and the time-dependent part of the total average  
force $\langle F\rangle=\langle F_l\rangle+\langle F_r\rangle$ does not
vanish. However, if in addition to the ensemble average  
one also evaluates the
time average over the period of modulation,  
the function 
$\xi_\alpha^{-2}(t)\,\cos\omega_\alpha t$ vanishes, and so do the 
total average force and the average velocity of the particle.
The absence of the drift in this case can also be directly demonstrated solving
the Langevin equation ({\ref{LE_V}).  
Thus  the linear response approximation (\ref{F_linear}) is insufficient to  account for 
the operation  of the system as a motor.

The last two terms in the right hand side of Eq. (\ref{F_alpha}) are of second
order in $\omega_\alpha$ and present the correction to the linear respond
approximation.  Both terms do not vanish under  time averaging 
over the modulation period and thus can generate a drift of  the particle. 
As one can check, the time averages  of these two terms  are of  
opposite signs.

We finish this section with a remark that our derivation of Eq. 
(\ref{F_alpha}) for the average forces 
does not assume that the modulation amplitudes $a_\alpha$
are small. An additional assumption
$a_\alpha\ll 1$ and the linear approximation  
for the functions $\xi_\alpha^{-n}(t)$ 
\begin{eqnarray}
\xi_\alpha^{-n}(t)\approx1-a_\alpha\,n\,\sin\omega_\alpha t
\end{eqnarray}
bring Eq. (\ref{F_alpha}) to order $a_\alpha^2$ to the following form 
\begin{eqnarray}
\langle F_\alpha(t)\rangle&=&\pm\, F_0\,\Bigl\{\,1+ 
c_1\,\,(\omega_\alpha\hat\tau_{\alpha})\,\,a_\alpha\,\, 
\cos\omega_\alpha t-c_1\,\,(\omega_\alpha\hat\tau_{\alpha})\,\,a_\alpha^2\,\, 
\sin 2\,\omega_\alpha t\nonumber\\
&+&c_2\,\,(\omega_\alpha\hat\tau_{\alpha})^2\,\,a_\alpha\,\, 
\sin\omega_\alpha t+(\omega_\alpha\hat\tau_{\alpha})^2\,\,a_\alpha^2\,\, 
[c_3\,\cos^2\omega_\alpha t-3\,c_2\,\sin^2\omega_\alpha t] \,
\Bigr\}.\nonumber
\label{F_alpha_small_a}
\end{eqnarray}
Clearly, only the last term, quadratic in both $\omega_\alpha$ and $a_\alpha$,
contributes to the drift of the motor.  Thus for the present model the drift
is a nonlinear phenomenon with respect to both modulation parameters. 

When both $\omega_\alpha$ and $a_\alpha$ are small the drift is small too and
numerical simulation becomes very time-consuming.  
In remaining sections we shall work with the  more general expression 
(\ref{F_alpha}) which does not assume the smallness of $a_\alpha$.

\section{Results}
In this section we discuss solutions of the Langevin equation (\ref{LE_V})
with  the average fluctuating force given by Eq. (\ref{F_alpha}).  
Although the equation is linear, its explicit analytic
solutions are rather bulky and not instructive. We therefore present 
solutions in a graphical form only.

Let us first consider the case already discussed  in Section II 
when the asymmetry is due to unequal magnitudes of the passive
forces $f_l\ne f_r$, 
while the modulation 
frequencies and amplitudes for the left and right sides are the same,
\begin{eqnarray}
\!\!\!\!
\omega_l=\omega_r=\omega, \quad a_l=a_r=a, \quad \xi_l(t)=\xi_r(t)=\xi(t). 
\end{eqnarray}
Introducing the parameter of asymmetry $\delta$
by relations
\begin{eqnarray}
f_r=\delta \,f_l, \quad \mbox{or}\quad \hat\tau_r=\delta^{-1}\,\,\hat\tau_l,
\end{eqnarray}
one obtains from (\ref{F_alpha}) the following expression
for  the total average  force 
$\langle F\rangle= \langle F_l\rangle+\langle F_r\rangle$:
\begin{eqnarray}
\!\!\!\!\!\!\!\!\!\!\!\!\!\!
\langle F(t)\rangle&=&F_0\,\Bigl\{
c_1\,(\omega\hat\tau_{l})\,\left(1-\delta^{-1}\right)\,a\,\xi^{-2}(t)\, 
\cos\omega t\nonumber\\
&+&c_2\,(\omega\hat\tau_{l})^2\,
\left(1-\delta^{-2}\right) \,a\,\xi^{-3}(t)\, 
\sin\omega t\nonumber\\
&+&c_3\,(\omega\hat\tau_{l})^2\,
\left(1-\delta^{-2}\right)\,a^2\,\xi^{-4}(t)\,\cos^2\omega t\,
\Bigr\}.
\label{F_A}
\end{eqnarray}
This expression is to be substituted into 
the Langevin equation (\ref{LE_V}), which we integrate numerically 
using the following time, space, 
and velocity units
\begin{eqnarray}
t_0=\frac{1}{2}\,\hat\tau_l=\frac{mv_T}{f_l},\quad 
x_0=v_T\,t_0, \quad
v_0=\frac{x_0}{t_0}=v_T.
\label{units}
\end{eqnarray}
The same units of course are  employed in simulation. (Note that the
acceleration unit is $a_0=v_0/t_0=f_l/m$, which means that 
molecules in the left interaction zone have acceleration of magnitude one).

\begin{figure}[htb]
\centerline{\includegraphics[scale=0.65]{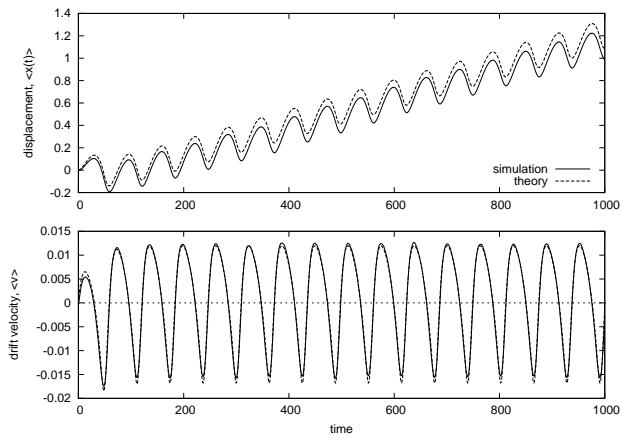}}
\caption{Average displacement $\langle x(t)\rangle$ and
velocity  $\langle v(t)\rangle$ of the motor as functions of time for 
regime (\ref{D0}) with parameters  
$\omega=0.1$, $a=0.5$, and $f_r=f_l=1$.
Simulation  data (solid lines) are averaged over about 
$2\cdot 10^5$ trajectories. Theoretical curves (dashed lines) are solutions
of the Langevin equation (\ref{LE_dimless_2}).    
}
\label{fig7}
\end{figure}   

In dimensionless form the Langevin equation (\ref{LE_V}) with 
the force (\ref{F_A}) and $\gamma$ given by (7) reads as follows
\begin{eqnarray}
\!\!\!\!\!\!\!\!\!\!\!\!\!\!
\frac{d}{d\tilde t}\, \langle \tilde V\rangle=&-&3\,c_1\,\lambda^2\,\tilde n \,\langle
\tilde V\rangle+2\,c_1\,\lambda^2\,\tilde n\,a\,\tilde\omega\,
(1-\delta^{-1})\,\,\xi^{-2}(t)\,\,\cos\tilde\omega\tilde t \nonumber\\ 
&+&4\,c_2\,\lambda^2\,\tilde n\, a\,\tilde\omega^2\,
(1-\delta^{-2})\,\,\xi^{-3}(t)\,\,\sin\tilde\omega\tilde t\nonumber\\ 
&+&4\,c_3\,\lambda^2\,\tilde n\, a^2\,\tilde\omega^2\,
(1-\delta^{-2})\,\,\xi^{-4}(t)\,\,\cos^2\tilde\omega\tilde t, 
\label{LE_dimless_1}
\end{eqnarray} 
where $\lambda=\sqrt{m/M}$, and a superposed  
tilde is used to denote dimensionless velocity 
$\tilde V=V/v_0$, time $\tilde t=t/t_0$, 
frequency $\tilde\omega=\omega\, t_0$, and 
the concentration of bath molecules $\tilde n=n\,x_0$. 
Since the operations of taking a time derivative $d/dt$ and an ensemble average 
$\langle ...\rangle$ clearly commute, the equation
for the dimensionless displacement $\tilde X=X/x_0$ follows 
from (\ref{LE_dimless_1}) simply by the replacement
$\langle \tilde V\rangle =\frac{d}{d\tilde t}\langle \tilde X\rangle$.   
The solutions for $\langle \tilde X(\tilde t)\rangle $ and 
$\langle \tilde V(\tilde t)\rangle $
are presented in Fig. 2 for zero initial conditions by dashed lines. 
They are in good agreement with numerical simulation (solid lines)
provided the modulation amplitude is not too high, $a\le 0.5$.

Motors with other types of asymmetry can be considered in similar ways.
As another example, let us compare predictions of the theory  
and simulation results for the case 
when the left side of the particle is active, while 
the right side is passive, 
\begin{eqnarray}
\omega_l=\omega\ne 0, \quad a_l=a\ne 0\qquad \omega_r=a_r=0.
\label{D0}
\end{eqnarray}
For this case, the general
expression (\ref{F_alpha})  gives the following result for the 
ensemble-average
fluctuating force on the particle
\begin{eqnarray}
\!\!\!\!\!\!\!\!\!\!\!\!\!\!
\langle F(t)\rangle&=&F_0\,\Bigl\{
c_1\,(\omega\hat\tau)\,a\,\xi^{-2}(t)\, \cos\omega t\nonumber\\
&+&c_2\,(\omega\hat\tau)^2\,a\,\xi^{-3}(t)\, \sin\omega t
+c_3\,(\omega\hat\tau)^2\,a^2\,\xi^{-4}(t)\,\cos^2\omega t\,
\Bigr\},
\label{F_D}
\end{eqnarray}
where all quantities are for the particle's left side. 
The corresponding Langevin  equation for the average velocity 
in dimensionless notations
has the form 
\begin{eqnarray}
\!\!\!\!\!\!\!\!\!\!\!\!\!\!\!
\frac{d}{d\tilde t}\, \langle \tilde V\rangle=&-&3\,c_1\,\lambda^2\,\tilde n \,\langle
\tilde V\rangle
+2\,c_1\,\lambda^2\,\tilde n\,a\,\tilde\omega\,\xi^{-2}(t)\,\,\cos\tilde\omega\tilde t 
\nonumber\\ 
&+&4\,c_2\,\lambda^2\,\tilde
n\,a\,\tilde\omega^2\,\xi^{-3}(t)\,\,\sin\tilde\omega\tilde t
+4\,c_3\,\lambda^2\,\tilde n\,a^2\,\tilde\omega^2\,\xi^{-4}(t)\,\,\cos^2\tilde\omega\tilde t. 
\label{LE_dimless_2}
\end{eqnarray}

The solution of this equation and of  a similar equation for the
dimensionless displacement $\langle\tilde X(\tilde t)\rangle$
are presented by dashed lines in Fig. 7. 
Again, within the range of its
validity, $\omega\hat\tau\ll 1$,
the theory agrees with simulation provided $a\le 0.5$. As 
the variation amplitude $a$ is further increased, 
the theory progressively overestimates the drift velocity.  
We believe this is due to neglected 
effects of the variation 
on the dissipative force in the Langevin equation.
A small offset between theoretical and experimental curves
visible in both Fig. 2 and Fig. 7 can 
probably be attributed to non-Markovian effects which  
are also neglected in the presented theory.
We hope to address these issues  in future work.

\section{Conclusion}
The relations between internal conformational dynamics of complex 
biomolecular systems and their ability to move directionally are of 
considerable interest in many fields. 
Ab initio simulation of coupled conformational and translational dynamics
is usually of high computational cost and often not illuminating.  
In this paper we studied a reduced model
where the internal dynamics of a motor is not considered explicitly 
but is assumed to result in periodic variation of the strength of the 
microscopic potential through which the motor interacts with 
molecules of the surrounding thermal bath. The model is simple enough to
allow the evaluation of the driving force in an analytic form. 
In contrast to many models with variation of external parameters,
our model shows  no directional motion in adiabatic approximation
of infinitely slow (reversible) variations. Moreover, the leading-order 
correction to the adiabatic approximation 
(linear in the variation frequency $\omega$)
is also insufficient: Without the last two terms quadratic in $\omega$,
the Langevin equation (\ref{LE_dimless_2}), 
or (\ref{LE_dimless_1}), 
has a no-drift solution. Interestingly, those two terms, if only one of them
is left in (\ref{LE_dimless_2}), generate the drift in opposite directions.
Thus, being the interplay of two terms, the direction of motion 
of the motor can hardly be predicted with  qualitative arguments.

The model can be extended in many ways. Since the role of internal variations
is to drive the motor out of equilibrium, it is clear that
a strictly periodic and deterministic character of variations is 
unnecessary to maintain the drift. One may expect a similar mechanism 
of directional transport, for instance,  for motors driven
by internal conformational transitions which, due to the fuel consumption, 
do not satisfy the equilibrium detailed balance condition. 
Models with stochastic transitions between two and more internal 
conformational states were discussed earlier~\cite{Prost}, 
but they implied the existence of the external potential(s), whereas 
our motor is autonomous.

\begin{acknowledgements}
I thank to G. Buck, G. Parodi, and J. Schnick for discussions, and to 
an anonymous referee for valuable comments.

\end{acknowledgements}

\section*{Appendix}
This Appendix (not included in the published version) presents
perturbation solutions of Eq. (41) to second
order in the small parameter $\epsilon=\omega\tau^0$. 
The equation has a form
\begin{eqnarray}
1+c_1\,x+ \epsilon\, c_2 \,x^2+\epsilon^2\, c_3 \,x^3=0
\nonumber
\end{eqnarray}
where $x=\tau^*/\tau^0$ and 
\begin{eqnarray}
c_1=-\xi(t),\quad
c_2=-\frac{a}{3}\,\cos\omega t,\quad
c_3=\frac{a}{12}\,\sin\omega t.
\nonumber
\end{eqnarray} 
Substituting into the equation the second-order ansatz 
\begin{eqnarray}
x=x_0+\epsilon\,x_1+\epsilon^2\,x_2
\nonumber
\end{eqnarray}
and discarding terms of order
higher than two yields
\begin{eqnarray}
1+c_1 (x_0+\epsilon\, x_1+\epsilon^2\, x_2)+
\epsilon\, c_2\, (x_0^2+2\,\epsilon\, x_0\, x_1)
+\epsilon^2\, c_3\, x_0^3=0.
\nonumber
\end{eqnarray}
Equating to zero contributions of each order separately
\begin{eqnarray}
1+c_1\,x_0=0,\quad
c_1\,x_1+c_2\,x_0^2=0,\quad
c_1x_2+2\,c_2\,x_0\,x_1+c_3\,x_0^3=0,
\nonumber
\end{eqnarray}
one finds
\begin{eqnarray}
x_0=-\frac{1}{c_1}, \quad
x_1=-\frac{c_2}{c_1^3},\quad
x_2=\frac{c_3}{c_1^4}-\frac{2\,c_2^2}{c_1^5}.
\nonumber
\end{eqnarray}
For $\tau^*=x\,\tau^0$ these relations lead
to the solution  (\ref{tau_star}). One can show that 
two other solutions 
are singular (diverge as $\epsilon\to 0$) and have no physical meaning.


\begin{thebibliography}{99}






\bibitem{Kampen} N. Van Kampen, 
{\it Stochastic Processes in Physics and Chemistry}, North-Holland, 2007,
Chapter V. 

\bibitem{granular} B. Cleuren and C. Van den Broeck, 
{\it Granular Brownian motor},
Europhys. Lett. 77, 50003 (2007); 
J. Talbot, R. D. Wildman, and P. Viot, 
{\it Kinetics of a Frictional Granular Motor},
Phys. Rev. Lett. 107, 138001 (2011);
A. Sarracino, A. Gnoli, and A. Puglisi, 
{\it Ratchet effect driven by Coulomb friction: The asymmetric Rayleigh piston},
Phys. Rev. E 87, 040101(R) (2013). 



\bibitem{chem} R. Golestanian, T. B. Liverpool, and A. Ajdari,
{\it Propulsion of a molecular machine by asymmetric distribution 
of reaction products},
Phys. Rev. Lett. 94, 220801 (2005);
G. R\"uckner and R. Kapral,
{\it Chemically Powered Nanodimers},
Phys. Rev. Lett. 98, 150603 (2007);
P. H. Colberg and R. Kapral,
{\it Angstrom-scale chemically powered motors},  EPL 106 30004 (2014).



\bibitem{info} D. Abreu and U. Seifert, 
{\it Extracting work from a single heat bath through feedback},
EPL 94, 10001 (2011);
D. Mandal and Ch. Jarzynski,   
{\it 
Work and information processing in a solvable model of Maxwell’s demon},
PNAS 109, 11641  (2012);
Z. Lu, D. Mandal, and Ch. Jarzynski, 
{\it Engineering Maxwell’s demon},
Phys. Today 67, 60 (2014).



\bibitem{Mecke}
S. Sporer, Ch. Goll, and K. Mecke, 
{\it 
Motion by stopping: Rectifying Brownian motion of nonspherical particles},
Phys. Rev. E 78, 011917 (2008). 


\bibitem{Reimann_rev1}
P. Reimann and P. H\"anggi,  
{\it Introduction to the physics of Brownian motors},
Appl. Phys A 75, 169 (2002).

\bibitem{Reimann_rev2}
P. Reimann,  
{\it Brownian motors: noisy transport far from equilibrium},
Phys. Rep, 361,  57  (2002).


\bibitem{activeBM} 
P. Romanchzuk, M. B\"ar, W. Ebeling, B. Lindner,
 and L. Schimancky-Geier, 
{\it Active Particles: From Individual to Collective Stochastic Dynamics},
Eur. Phys. J. Special-Topics 202, 1 (2012).



\bibitem{Gruber} 
Ch. Gruber and J. Piasecki, {\it
Stationary motion of the adiabatic piston},   
Physica A 268, 412 (1999).

\bibitem{Broeck_piston}
E. Kestemont, C. Van den Broeck, and M. Malek Mansour, 
{\it The “adiabatic” piston: And yet it moves},
Europhys. Lett, 49, 
143 (2000).

\bibitem{Munakata}
T. Munakata and H. Ogawa, 
{\it Dynamical aspects of an adiabatic piston},
Phys. Rev. E 64, 036119 (2001)


\bibitem{Broeck_motor1} 
C. Van den Broeck, R. Kawai, and P. Meurs,
{\it Microscopic analysis of a thermal Brownian motor},
Phys. Rev. Lett. 93, 090601 (2004). 

\bibitem{Broeck_motor2}
P. Meurs, C. Van den Broeck, and A. Garcia,  
{\it Rectification of thermal fluctuations in ideal gases}, Phys. Rev. E 
70, 051109 (2004).

\bibitem{Broeck_motor3}
P. Meurs and C. Van den Broeck, 
{\it Thermal Brownian motor},
J. Phys.: Condens. Matter 17, S3673 (2005).

\bibitem{Plyukhin_Froese}
A. V. Plyukhin and A. M. Froese, 
{\it Nonlinear dissipation effect in Brownian relaxation},
Phys. Rev. E 76, 031121 (2007).

\bibitem{Plyukhin_piston} 
A. V. Plyukhin and J. Schofield, 
{\it Langevin equation for the extended Rayleigh model with 
an asymmetric bath}, Phys. Rev. E 69, 021112 (2004).



\bibitem{Reimann_motor}
P. Reimann, R. Bartussek, R. H\"aussler, and P. H\"anggi, 
{\it Brownian Motors Driven by Temperature Oscillations},
Phys. Lett. A 215, 26 (1996).




\bibitem{Reimann_motor2} 
S. von Gehlen, M. Evstigneev, and P. Reimann,
{\it Ratchet effect of a dimer with broken friction 
symmetry in a symmetric potential},
Phys. Rev. E 79, 031114 (2009).


\bibitem{Parrondo} J. M. R. Parrondo, 
{\it 
Reversible ratchets as Brownian particles in an adiabatically 
changing periodic potential},
Phys. Rev. E 57, 7297 (1998).






\bibitem{Astumian} 
R. D. Astumian, 
{\it Adiabatic operation of a molecular machine},
Proc. Natl. Acad. Sci. 104, 19715 (2007).


\bibitem{Jar} 
S. Rahav, J. Horowitz, and Ch. Jarzynski, 
{\it Directed Flow in Nonadiabatic Stochastic Pumps},
 Phys. Rev. Lett. 101, 140602 (2008).

\bibitem{Cher} 
V. Y. Chernyak and  N. A. Sinitsyn, 
{\it Pumping Restriction Theorem for Stochastic Networks},
Phys. Rev. Lett. 101, 160601 (2008). 


\bibitem{Plyukhin_motor} 
A. V. Plyukhin, {\it 
Brownian diode: Molecular motor based on a semi-permeable Brownian 
particle with internal potential drop},
Phys. Lett. A 377, 1037 (2013).

\bibitem{Prost}
J. Prost, J.-F. Chauwin, L. Peliti, A. Ajdari,
{\it Asymmetric pumping of particles}, 72, 2652 (1994);
F. J\"ulicher, A. Ajdari, J. Prost, 
{\it Modeling molecular motors}, Rev. Mod. Phys. 69, 1269 (1997);
H. Hagman, M. Zelan, C. M. Dion, 
{\it Breaking the symmetry of a Brownian motor with symmetric potentials}, 
J. Phys. A 44 (15), 155002 (2011).


\end{thebibliography}
\end{document}